\documentclass[a4paper,12pt]{article}
\usepackage{amsmath}
\usepackage{amsfonts}
\usepackage{mathrsfs}
\usepackage{amsthm}
\usepackage{extarrows}
\usepackage{bm}
\usepackage{graphicx}
\usepackage[section]{placeins}
\usepackage{flafter}
\usepackage{array}
\usepackage{caption}
\usepackage{subcaption}
\usepackage{color}
\usepackage{multirow}
\usepackage{natbib}
\usepackage{enumerate,url}
\usepackage{algorithmic}
\usepackage{algorithm}

\newtheorem{thm}{Theorem}

\usepackage{xr}
\usepackage[small,compact]{titlesec}

\title{Parallel subgroup analysis of high-dimensional data via M-regression}
\author{Chao Cheng and Xingdong Feng \footnote{\textit{Address for correspondence:} School of Statistics and Management \& Institute of Data Science and Statistics, Shanghai University of Finance and Economics, 777 Guoding Road, Shanghai 200433, China. Email: feng.xingdong@mail.sufe.edu.cn. This work  was supported by National Natural Science Foundation of China (11971292 and 11690012), and Program for Innovative Research Team of SUFE.}\\
\textit{Shanghai University of Finance and Economics}}
\date{\today}

\begin{document}
\maketitle

\begin{abstract}
  It becomes an interesting problem to identify subgroup structures in data analysis as populations are probably heterogeneous in practice. In this paper, we consider M-estimators together with both concave and pairwise fusion penalties, which can deal with high-dimensional data containing some outliers. The penalties are applied both on covariates and treatment effects, where the estimation is expected to achieve both variable selection and data clustering simultaneously.  An algorithm is proposed to process relatively large datasets based on parallel computing. We establish the convergence analysis of the proposed algorithm, the oracle property of the penalized M-estimators, and the selection consistency of the proposed criterion. Our numerical study demonstrates that the proposed method is promising to efficiently identify subgroups hidden in high-dimensional data.
\end{abstract}

\clearpage
\section{Introduction}
\label{sec:introduction}
Heterogeneity is often present in practice, which raises many challenges, but significant improvements can be made on  data analysis if it is successfully taken into account. For example, in precision medicines and individualized treatment designs, it is widely recognized that  treatment effects may vary among different patients. Hence it is important to precisely characterize the personal attributes for a better prescription. Typical subgroup analysis in a clinical trial splits the studied population into subgroups based on observed features, such as age, gender, race and so on. For identifying latent heterogeneity, it is popular to consider  data as coming from mixture models \citep[among others]{Everitt2011p-, Banfield1993p803-803, Hastie1996p155-176, McNicholas2010p1175-1181, Shen2015p303-312}. Mixture models can easily incorporate covariate effects due to its theoretical framework, but the number of subgroups and the corresponding underlying distributions are usually pre-specified, which is hard to be verified.
\par
Penalized methods can identify hidden subgroups and estimate their centers simultaneously, without any prior information of true subgroup structures. Penalization for unsupervised clustering has been consistently considered by \cite{Hocking2011p745-752}, \cite{Chi2015p994-1013}, \cite{Pan2013p1865-1889} and \cite{Wu2016p1-25}. \cite{Ma2017p410-423}  not only utilize the pairwise fusion penalty among all the subjects to detect the hidden subgroups, but also use the mean regression  to incorporate the covariate effects. \cite{ZHANG2019p-} consider the median regression which is more robust to outliers. Since most of these methods are based on over-parameterization of the centroids, algorithms designed to solve these problems often own the complexity at the quadratic order of the sample size, which limits the application to  large-scale  data. To address this issue, one possible solution is parallel or distributed computation, such as the divide-and-conquer strategy mentioned by \cite{ZHANG2019p-}.
\par
In this paper, we equip the regression-based subgroup analysis with certain variable selection methods to choose useful variables from a large number of candidate covariates, which is less touched in previous penalized-regression clustering literature. Variable selection has always been vital in high-dimensional statistical analysis, especially when  numbers of observations and covariates are both exploding. Suitable variable selection methods can improve interpretability of a model and  understanding of data. In this paper, we have established the oracle properties for both subgroup recovery and variable selection with a variety of penalties, such as  SCAD \citep{fan1} and MCP \citep{zhang}. We consider M-estimators \citep{huber}, which include the least squares estimator of \cite{Ma2017p410-423} and the least absolute deviation estimator of \cite{ZHANG2019p-} as the special cases. 
\par
The alternating direction method of multipliers (ADMM) is a simple yet powerful optimization method for parallel and distributed computations on large-scale data. It is first introduced by \cite{Gabay1976p17-40} and \cite{Glowinski1975p327-333}. The convergence properties of ADMM on convex problems are established by \cite{Glowinski1984p-} and further studied by \cite{Davis2016p115-163}, \cite{Davis2017p783-805} and \cite{He2012p700-709}. We refer to  \cite{Boyd2010p1-122} for a comprehensive review of this method. Although in general ADMM may fail as  nonconvex objective functions are considered,  it has been found to perform well in some interesting cases \citep{Hong2016p337-364, Li2015p2434-2460, Wang2018p-}. In this paper, we propose an ADMM-based algorithm, and
implement it in a parallel manner to speed up its computation as we analyze the relatively large data. It is worthy of pointing out that the number of pairwise penalties over individualized treatment effects increases at the quadratic order of the sample size, which makes it the bottleneck of almost every penalized regression clustering algorithm. We facilitate this problem by considering parallel computation by elementwisely updating augmented variables. Furthermore, we observe the phenomenon that the ADMM algorithm can provide a reasonable result after some iterations quite quickly, although it would take much more iterations to meet a more strict convergence criterion, which is also mentioned by \cite{Boyd2010p1-122}. We take this characteristic of the ADMM algorithm into account, and in practice  we could stop the ADMM algorithm in an early stage, and use a simple clustering method, such as the K-means method to further improve grouping results. This strategy is practically effective, which is demonstrated in our numerical study.
\par
The rest of this paper is organized as follows. In Section~\ref{sec:model-setting} we introduce the estimation procedure. In Section~\ref{sec:proposed-algorithm} we propose the parallel algorithm together with its convergence analysis. In Section~\ref{sec:asympt-properties} we discuss the theoretical properties of the penalized M-estimator. In Section~\ref{sec:simulation-result} we assess the performance of the proposed method  under different settings in a simulation study. The detailed proofs of Theorems \ref{sec:thm:convergence-of-the-algorithm}--\ref{thm:consistency_of_mBIC}  are deferred to the Supplementary Material. An R package implementing the proposed method is available at \url{https://github.com/fenguoerbian/RSAVS}.

\section{Model Setting}
\label{sec:model-setting}
In this paper, we consider the following model
\[
  y_i = \mu_i + \bm{x}_i^T\bm{\beta} + \varepsilon_i, \quad\quad i = 1, 2, \cdots, n,
\]
where $\bm{x}_i$ is  the $p$-dimensional covariate of the $i$-th subject and $\bm{\beta}$ is the corresponding coefficient, $\varepsilon_i$'s are independently distributed  errors, and $\mu_i$ is the individual treatment effect attributable to the $i$-th subject. The observations are  collected from the unknown $K_0$ different subgroups, with the true intercepts $\bm{\alpha}_0 = \left(\alpha_{0, 1}, \cdots, \alpha_{0, K_0}\right)^T$. These groups are denoted by $G_1, \cdots, G_{K_0}$ and $i\in G_k$ if the $i$-th observation belongs to the $k$-th subgroup, which implies that
\[
  \mu_i = \alpha_{0, k},\quad\quad \text{if }i \in G_k.
\]
Then the model can also be written as
\[
  y_i = \bm{z}_i^T\bm{\alpha} + \bm{x}_i^T\bm{\beta} + \varepsilon_i,
\]
where $\bm{z}_i$'s are the $K_0$-dimensional indicator vectors, and $z_{ij} = 1$ if $i\in G_j$ and $z_{ij} = 0$ if $i\notin G_j$. The length of the coefficient $\bm{\beta}$ is $p$, which can grow even faster than the sample size $n$. We assume the sparsity structure is present, which means only a small portion of the elements of the vector $\bm{\beta}$ is non-zero. Mathematically, suppose that only $q$ entries of the coefficient $\bm{\beta}$ are non-zero while the rest are zeros. For easy presentation, we write $\bm{\beta}_0 = \left(\bm{\beta}_{A}^T, \bm{0}_{p - q}^T\right)^T$, and accordingly, the covariate vector is given as $\bm{x}_i = \left(\bm{x}_{A,i}^T, \bm{x}_{I, i}^T\right)^T$, where $A$ and $I$ refer to the active and inactive sets of predictor indices respectively. Note that $q$ can also increase as $n$ grows, but with a much slower rate. Here we need to point out that the sets $A$ and $I$ are unknown.
\par
Unfortunately, in practice the real parameter structure is often unknown, which includes the grouping information and the set of active coefficients. We define the proposed penalized estimator of $\left(\bm{\mu}, \bm{\beta}\right)$ as
\begin{equation}
  \label{eq:penalized_objective_function}
  \left(\hat{\bm{\mu}}, \hat{\bm{\beta}}\right) =
  \underset{\bm{\mu}, \bm{\beta}}{\mathrm{arg\,min}}\;
  \frac{1}{n}\sum\limits_{i = 1}^n\rho\left(y_i - \mu_i - \bm{x}_i^T\bm{\beta}\right)
  + \sum\limits_{1 \leq i < j \leq n}P_{\lambda_1}\left(\mu_i - \mu_j\right)
  + \sum\limits_{j = 1}^pP_{\lambda_2}\left(\beta_j\right),
\end{equation}
where the first part on the right hand side is just a regular loss function $\rho$ of M-regression, and the second and third terms are used to identify the subgroup structures and the active coefficients, respectively.   Possible choices of the loss function $\rho$ include the least squares $L_2$, the absolute deviance $L_1$ and the Huber loss \citep{huber}, denoted as $Huber$. With the $L_2$ loss, the solution of (\ref{eq:penalized_objective_function}) will be easier to obtain. With the $L_1$ or the $Huber$ losses, the results are expected to be more robust to outliers.
\par
In this paper, we assume that $\mathrm{E}\psi\left(\varepsilon_i\right) = 0$, where $\psi$ is the derivative (or directional derivative) of $\rho$, and we allow $\rho$ to be differentiable except at finitely many points. Regarding the penalties, the popular choices, such as SCAD and MCP, have been widely used and well studied in the literature. These penalties can shrink some pairs $\mu_i -\mu_j$ and coefficients $\beta_j$ to be zeros, which gives us an estimated parameter structure based on the data.
\par
Although the penalties SCAD and MCP are nonconvex, they are continuous and even. For the SCAD method, we consider the following penalty function
\[
  P'_{\lambda, \gamma}\left(x\right) =
  \lambda\left\{
    I\left(x \leq \lambda\right)
    + \frac{\left(\gamma\lambda - x\right)_+}{\left(\gamma - 1\right)\lambda}
    I\left(x > \lambda\right)
  \right\},\quad\quad x > 0,\quad \gamma > 2,
\]
and for the MCP method, we use the following penalty function
\[
  P'_{\lambda, \gamma}\left(x\right) = \lambda
  \left(1 - \frac{x}{\lambda\gamma}\right)_+,\quad\quad x > 0,\quad \gamma > 1.
\]
\par
We use $S_0$ to represent the true parameter structure. The oracle estimator $\left(\tilde{\bm{\mu}}, \tilde{\bm{\beta}}\right)$is obtained under the assumption that $S_0$ is known, which is given by
\begin{equation}
  \label{eq:oracle_objective_function}
  \left(\tilde{\bm{\alpha}}, \tilde{\bm{\beta}}\right)
  = \left(\tilde{\bm{\alpha}}, \left(\tilde{\bm{\beta}}_A^T, \bm{0}^T_{p-q}\right)^T\right)
  = \underset{\bm{\alpha}, \bm{\beta}_A}{\mathrm{arg\,min}}\;
  \frac{1}{n}\sum\limits_{i = 1}^n\rho\left(
    y_i - \bm{z}_i^T\bm{\alpha} - \bm{x}_{A,i}^T\bm{\beta}_A
  \right),
\end{equation}
and $\tilde{\mu}_i = \tilde{\alpha}_k$ if $i \in G_k$. The proposed estimator is obtained by solving an nonconvex optimization problem, which will be further discussed in the following section.

\section{Proposed Algorithm}
\label{sec:proposed-algorithm}
The ADMM is widely used in statistical learning and optimization problems, especially as large-scale data are analyzed. This method changes a large optimization problem into smaller ones and it takes advantages of the augmented Langrangian method and the coordinate descent method. First we rewrite our original objective function in a new form
\begin{equation}
  \label{eq:objective_function_in_optimization_form}
  \begin{aligned}
    & \mathrm{min}\;
    \frac{1}{n}\sum\limits_{i = 1}^n \rho\left(z_i\right)
    + \sum\limits_{1 \leq i < j \leq n}P_{\lambda_1}\left(s_{ij}\right)
    + \sum\limits_{j = 1}^pP_{\lambda_2}\left(w_j\right)    \\
    & \mathrm{s.t.}\;
    \left\{
      \begin{aligned}
        & \bm{z} = \bm{y} - \bm{\mu} - \bm{X}\bm{\beta}    \\
        & \bm{s} = \bm{D}\bm{\mu}    \\
        & \bm{w} = \bm{\beta}
      \end{aligned},
    \right.
  \end{aligned}
\end{equation}
where $\bm{D}$ is the $\frac{n\left(n - 1\right)}{2} \times n$ pairwise difference matrix and $s_{ij} = \mu_i - \mu_j$. We use double subscript in $s_{ij}$ to refer to the couple differences between those individual treatment effects. And the augmented Lagrangian form of this problem is then given by
\begin{equation}
  \label{eq:augmented_lagrangian_objective_function}
  \begin{aligned}
    & L\left(\bm{\beta}, \bm{\mu}, \bm{z}, \bm{s}, \bm{w}, \bm{q}_1, \bm{q}_2, \bm{q}_3\right)    \\
    =& \frac{1}{n}\sum\limits_{i = 1}^n\rho\left(z_i\right)
    + \sum\limits_{1\leq i < j\leq n}P_{\lambda_1}\left(s_{ij}\right)
    + \sum\limits_{j = 1}^pP_{\lambda_2}\left(w_j\right)    \\
    +& \frac{r_1}{2}\left\|\bm{y} - \bm{\mu} - \bm{X}\bm{\beta} - \bm{z} \right\|_2^2
    + \frac{r_2}{2}\left\|\bm{D}\bm{\mu} - \bm{s}\right\|_2^2
    + \frac{r_3}{2}\left\|\bm{\beta} - \bm{w}\right\|_2^2    \\
    +& \left\langle \bm{y} - \bm{\mu} - \bm{X}\bm{\beta} - \bm{z}, \bm{q}_1 \right\rangle
    + \left\langle \bm{D}\bm{\mu} - \bm{s}, \bm{q}_2 \right\rangle
    + \left\langle \bm{\beta} - \bm{w}, \bm{q}_3 \right\rangle
    ,
  \end{aligned}
\end{equation}
where $r_1$, $r_2$ and $r_3$ are positive scalars and $\bm{q}_1$, $\bm{q}_2$ and $\bm{q}_3$ are multiplier vectors, and $\left\|\bm{a}\right\|_2^2 = \bm{a}^T\bm{a}$ and $\left\langle \bm{a}, \bm{b} \right\rangle = \bm{a}^T\bm{b}$.

We use an iterative algorithm to solve (\ref{eq:objective_function_in_optimization_form}) which is summarized in Algorithm \ref{alg:ADMM}. More details of this algorithm are provided in the following subsections.

\begin{algorithm}\footnotesize
  \caption{ADMM for subgroup analysis and variable selection}
  \label{alg:ADMM}
  \begin{algorithmic}[1]
    \REQUIRE Set and fix the values of $r_1$, $r_2$, $r_2$, $\lambda_1$, $\lambda_2$ and some other possible parameters needed in the penalty function, such as $\gamma$ in the SCAD method. Also set the initial values of $\bm{\beta}^{(0)}$, $\bm{\mu}^{(0)}$, $\bm{z}^{(0)}$, $\bm{s}^{(0)}$, $\bm{w}^{(0)}$, $\bm{q}_1^{(0)}$, $\bm{q}_2^{(0)}$ and $\bm{q}_3^{(0)}$. Let $ k = 0 $. Set the maximum iterations and the convergence tolerance  as $max\_iter$ and $tol$, respectively. Denote the primal residual at the $m$th iteration by
  \begin{equation}
    \label{eq:primal-residual-definition}
    \bm{r}^{\left(m\right)} =
    \begin{pmatrix}
      \bm{y} - \bm{u}^{\left(m\right)} - \bm{X}^T\bm{\beta}^{\left(m\right)} - \bm{z}^{\left(m\right)}    \\
      \bm{D}\bm{\mu}^{\left(m\right)} - \bm{s}^{\left(m\right)}    \\
      \bm{\beta}^{\left(m\right)} - \bm{w}^{\left(m\right)}
    \end{pmatrix},
  \end{equation}
  and the dual residual by
  \begin{equation}
    \label{eq:dual-residual-definition}
    \bm{\eta}^{\left(m + 1\right)} =
    \begin{pmatrix}
      r_1\left(\bm{z}^{\left(m + 1\right)} - \bm{z}^{\left(m\right)}\right)
      - r_2\bm{D}^T\left(\bm{s}^{\left(m + 1\right)} - \bm{s}^{\left(m\right)}\right)    \\
      r_1 \bm{X}^T\left(\bm{z}^{\left(m + 1\right)} - \bm{z}^{\left(m\right)}\right)
      - r_3\left(\bm{w}^{\left(m + 1\right)} - \bm{w}^{\left(m\right)}\right)
    \end{pmatrix},
  \end{equation}
  where $\bm{D}$ is given in (\ref{eq:objective_function_in_optimization_form}).
    \WHILE{$k < max\_iter$ \AND $\mathrm{max}\left(\bm{r}^{\left(k\right)}, \bm{\eta}^{\left(k\right)}\right) > tol$}
      \STATE Update $\bm{\beta}^{(k+1)}$ by
      $ \bm{\beta}^{(k+1)} = \underset{\bm{\beta}}{\mathrm{argmin}}\; L\left(\bm{\beta}, \bm{\mu}^{(k)}, \bm{z}^{(k)}, \bm{s}^{(k)}, \bm{w}^{(k)}, \bm{q}_1^{(k)}, \bm{q}_2^{(k)}, \bm{q}_3^{(k)} \right) $.
      \STATE Update $\bm{\mu}^{(k+1)}$ by
  $ \bm{\mu}^{(k+1)} = \underset{\bm{\mu}}{\mathrm{argmin}}\; L\left(\bm{\beta}^{(k+1)}, \bm{\mu}, \bm{z}^{(k)}, \bm{s}^{(k)}, \bm{w}^{(k)}, \bm{q}_1^{(k)}, \bm{q}_2^{(k)}, \bm{q}_3^{(k)} \right) $.
      \STATE Update $\bm{z}^{(k+1)}$ by
  $ \bm{z}^{(k+1)} = \underset{\bm{z}}{\mathrm{argmin}}\; L\left(\bm{\beta}^{(k+1)}, \bm{\mu}^{(k+1)}, \bm{z}, \bm{s}^{(k)}, \bm{w}^{(k)}, \bm{q}_1^{(k)}, \bm{q}_2^{(k)}, \bm{q}_3^{(k)} \right) $.
      \STATE Update $ \bm{s}^{(k+1)} $ by
  $ \bm{s}^{(k+1)} = \underset{\bm{s}}{\mathrm{argmin}}\; L\left(\bm{\beta}^{(k+1)}, \bm{\mu}^{(k+1)}, \bm{z}^{(k+1)}, \bm{s}, \bm{w}^{(k)}, \bm{q}_1^{(k)}, \bm{q}_2^{(k)}, \bm{q}_3^{(k)} \right) $.
      \STATE Update $ \bm{w}^{(k+1)} $ by
  $ \bm{w}^{(k+1)} = \underset{\bm{w}}{\mathrm{argmin}}\; L\left(\bm{\beta}^{(k+1)}, \bm{\mu}^{(k+1)}, \bm{z}^{(k+1)}, \bm{s}^{(k+1)}, \bm{w}, \bm{q}_1^{(k)}, \bm{q}_2^{(k)}, \bm{q}_3^{(k)} \right) $.
      \STATE Update $\bm{q}_1^{(k+1)}$, $\bm{q}_2^{(k+1)}$ and $\bm{q}_3^{(k+1)}$ by
  \[
    \left\{
      \begin{aligned}
        \bm{q}_1^{(k+1)} &= \bm{q}_1^{(k)}
        + r_1 \left( \bm{y}^{(k+1)} - \bm{\mu}^{(k+1)} - \bm{X} \bm{\beta}^{(k+1)} - \bm{z}^{(k+1)} \right)    \\
        \bm{q}_2^{(k+1)} &= \bm{q}_2^{(k)}
        + r_2 \left( \bm{D}\bm{\mu}^{(k+1)} - \bm{s}^{(k+1)} \right)    \\
        \bm{q}_3^{(k+1)} &= \bm{q}_3^{(k)}
        + r_3 \left( \bm{\beta}^{(k+1)} - \bm{w}^{(k+1)} \right)
      \end{aligned}.
    \right.
  \]
      \STATE $k \leftarrow k + 1$.
    \ENDWHILE
    \ENSURE $\bm{\beta}^{\left(k\right)}$, $\bm{\mu}^{\left(k\right)}$, $\bm{z}^{\left(k\right)}$, $\bm{s}^{\left(k\right)}$, $\bm{w}^{\left(k\right)}$.
  \end{algorithmic}
\end{algorithm}

\subsection{Updating details}
\label{sec:updates-variables}
In this subsection, we merely provide the details of the proposed algorithm. The derivations are given in Section \ref{sec:algorithm-details} of the Supplementary Material. We use the $L_1$ loss coupled with the SCAD penalty as the example, and the iterative steps for the $L_2$ and $Huber$ losses and other penalties are given in the Supplementary Material.
\begin{enumerate}[(I)]
\item
  Update of $\bm{\beta}^{\left(k+1\right)}$:
  if $p \leq n$ then
\[
  \bm{\beta}^{\left(k + 1\right)} =
  \left(r_1\bm{X}^T\bm{X} + r_3\bm{I}_p\right)^{-1}
  \left\{
    r_1\bm{X}^T\left(\bm{y} - \bm{\mu} - \bm{z}\right)
    + r_3\bm{w} + \bm{X}^T\bm{q}_1 - \bm{q}_3
  \right\},
\]
and if $ p > n $ then
\[
  \bm{\beta}^{\left(k+1\right)}
  = \frac{1}{r_3}\left\{
    \bm{I}_p - r_1 \bm{X}^T\left(r_1\bm{X}\bm{X}^T + r_3 \bm{I}_n\right)^{-1}\bm{X}
  \right\}\left\{
    r_1\bm{X}^T\left(\bm{y} - \bm{\mu} - \bm{z}\right)
    + r_3\bm{w} + \bm{X}^T\bm{q}_1 - \bm{q}_3
  \right\},
\]
which implies that the algorithm only needs to compute the inverse of a matrix with the size  $\mathrm{min}(n, p)$. This is very useful when the number of variables are much larger than the number of observations.

\item
  Update of $\bm{\mu}^{\left(k+1\right)}$:
  \[
  \bm{\mu} = \left(
    r_1 \bm{I}_n + r_2 \bm{D}^T\bm{D}
  \right)^{-1}
  \left\{
    r_1\left(
      \bm{y} - \bm{X}\bm{\beta} - \bm{z}
    \right)
    + r_2 \bm{D}^T\bm{s} + \bm{q}_1 - \bm{D}^T\bm{q}_2
  \right\}.
\]
\item
  Update of $\bm{z}^{\left(k+1\right)}$: here we use the $L_1$ loss as an example, and it can be computed elementwisely by
  \[
  z_i^{\left(k+1\right)} = S\left(
    y_i - \mu_i - \bm{x}_i^T\bm{\beta} + \frac{q_{1,i}}{r_1}, \frac{1}{nr_1}
  \right)
  ,\quad\quad i = 1,\cdots,n,
\]
where $S\left(x, \lambda\right)$ is the soft-thresholding function
\[
  S\left(x, \lambda\right) = \left\{
    \begin{aligned}
      & x - \lambda    &&\quad\quad \lambda < x    \\
      & 0    && \quad\quad \left|x\right| \leq \lambda    \\
      & x + \lambda    && \quad\quad x < -\lambda
    \end{aligned}.
  \right.
\]
However, when the loss is $L_2$, the variable $\bm{z}$ is no longer necessary and the algorithm is similar to that of \cite{Ma2017p410-423}, and we refer to Section \ref{sec:algorithm-with-l2} in the Supplementary Material for more details.

\item
  Update of $\bm{s}^{\left(k+1\right)}$: for the SCAD penalty with parameters $\lambda_1$ and $\gamma_1$, the update is given by
  \[
    s_{ij}^{\left(k+1\right)} = \left\{
    \begin{aligned}
      & S\left(\mu_i - \mu_j + \frac{q_{2, ij}}{r_2}, \frac{\lambda_1}{r_2}\right)
      && \quad\quad \left| \mu_i - \mu_j + \frac{q_{2,ij}}{r_2} \right| \leq \left(1 + \frac{1}{r_2}\right)\lambda_1    \\
      & \frac{
        S\left(\mu_i - \mu_j + \frac{q_{2,ij}}{r_2}, \frac{\gamma_1\lambda_1}{r_2\left(\gamma_1 - 1\right)}\right)
      }{1 - \frac{1}{r_2\left(\gamma_1 - 1\right)}}
      && \quad\quad  \left(1 + \frac{1}{r_2}\right)\lambda_1
      < \left| \mu_i - \mu_j + \frac{q_{2,ij}}{r_2} \right| \leq \gamma_1\lambda_1    \\
      & \mu_i - \mu_j + \frac{q_{2,ij}}{r_2}
      && \quad\quad \left| \mu_i - \mu_j + \frac{q_{2,ij}}{r_2} \right| > \gamma_1\lambda_1
    \end{aligned}
  \right. .
  \]
\item
  Update of $\bm{w}^{\left(k+1\right)}$: 
  for the SCAD penalty with parameters $\lambda_2$ and $\gamma_2$, the update is given by
  \[
    w_j^{\left(k+1\right)} = \left\{
    \begin{aligned}
      & S\left(\beta_j + \frac{q_{3, j}}{r_3}, \frac{\lambda_2}{r_3}\right)
      && \quad\quad \left| \beta_j + \frac{q_{3, j}}{r_3} \right| \leq \left( 1 + \frac{1}{r_3}\right)\lambda_2    \\
      & \frac{
        S\left(\beta_j + \frac{q_{3, j}}{r_3}, \frac{\gamma_2\lambda_2}{r_3\left(\gamma_2 - 1\right)}\right)
      }{1 - \frac{1}{r_3\left(\gamma_2 - 1\right)}}
      && \quad\quad \left( 1 + \frac{1}{r_3}\right)\lambda_2
      < \left| \beta_j + \frac{q_{3, j}}{r_3} \right| \leq \gamma_2\lambda_2    \\
      & \beta_j + \frac{q_{3, j}}{r_3}
      && \quad\quad \left|\beta_j + \frac{q_{3, j}}{r_3}\right| > \gamma_2\lambda_2
    \end{aligned}
  \right. .
\]
\end{enumerate}

\subsection{Convergence analysis}
\label{sec:conv-algor}
We next establish the convergence property of the proposed algorithm. Some conditions on the loss function $\rho$ are necessary.
\begin{enumerate}[({C}1)]
\item
  $\rho$ is continuous on $\mathcal{R}$, and differentiable almost everywhere except finite many points.

\item
  $\rho$ is convex.

\item
  $\rho$ has a unique minimal point at 0 and $\rho\left(0\right) = 0$.
\end{enumerate}
Conditions (C1)--(C3) allow a broad class of losses including $L_1$, $L_2$, and $Huber$. We summarize the convergence result in the following theorem.
\begin{thm}
  \label{sec:thm:convergence-of-the-algorithm}
    The residuals satisfy
  \[
    \underset{m \to \infty}{\mathrm{lim}}\, \left\|\bm{r}^{\left(m\right)}\right\|_2^2 = 0
    ,\quad\quad
    \underset{m \to \infty}{\mathrm{lim}}\, \left\|\bm{\eta}^{\left(m\right)}\right\|_2^2 = 0,
  \]
  for the Lasso, SCAD and MCP penalties under Conditions (C1)--(C3), where the primal residual $\bm{r}^{\left(m\right)}$ and the dual residual $\bm{\eta}^{\left(m\right)}$ are given in (\ref{eq:primal-residual-definition}) and (\ref{eq:dual-residual-definition}), respectively.
\end{thm}

This theorem shows that the proposed algorithm will converge to a stationary point since the residuals converge to zero. Detailed proof are deferred to Section \ref{sec:proof-conv-algor} of the Supplementary Material.

\subsection{Searching region for tuning parameters}
\label{sec:choice-lambdal-and-lambda2}
In this subsection, we give the upper bounds of the tuning parameters $\lambda_1$ and $\lambda_2$,  denoted as $\lambda_1^{\left(0\right)}$ and $\lambda_2^{\left(0\right)}$, respectively.
Clearly, for some large enough tuning parameters, all entries of $\bm{\mu}$ and $\bm{\beta}$ would be shrunk to a constant $c$ and zero respectively, where $c$ is determined by $\rho$. For the least squares loss $L_2$, this constant is the average of $\left\{y_i, i = 1, 2, \cdots, n\right\}$, while for the loss $L_1$, it is the median.

Now we provide a sufficient large choice of tuning parameters $\lambda_1^{\left(0\right)}$ and $\lambda_2^{\left(0\right)}$ as initial values for finding the solutoin path.
\[
  \lambda_1^{(0)} =
  \frac{1}{n}
  \left\| \bm{D}\left(\bm{D}^T\bm{D}\right)^{-1}
  \begin{pmatrix}
    \psi \left(y_{1} - c\right)    \\
    \psi \left(y_{2} - c\right)    \\
    \vdots    \\
    \psi \left(y_{n} - c\right)    \\
  \end{pmatrix}
  \right\|_{\infty},
\]
and
\[
  \lambda_2^{(0)} \geq \frac{\left\| \bm{d}_{1,p} \right\|_{\infty}}{n}
  = \frac{1}{n}
  \left\|
    \sum\limits_{i=1}^{n} \left(
      \psi \left(y_{i} - c\right)
    \right) \bm{x}_{i}
  \right\|_{\infty},
\]
where $\left(\bm{D}^T\bm{D}\right)^{-1}$ is the Moore-Penrose generalized inverse and the matrix $\bm{D}$ is given in (\ref{eq:objective_function_in_optimization_form}). 
These initial values lead to a homogeneous structure where all samples are assumed to be identically distributed.
Details are given in Section \ref{sec:algorithm-details} of the Supplementary Material.
\par

\subsection{Criterion for choosing tuning parameters}
\label{sec:bic-lambda}

We use a modified BIC to choose the tuning prameters $\lambda_1$ and $\lambda_2$, which is defined as
\begin{equation}
  \label{eq:modified_BIC}
  \mathrm{BIC}\left(\hat{\delta}\left(\bm{\lambda}\right)\right)
  = \mathrm{log}\left(
    \frac{1}{n}
    \sum\limits_{i = 1}^n\rho\left(
      y_i - \bm{z}_i^T\hat{\mu}_i - \bm{x}_i^T\hat{\bm{\beta}}
    \right)
  \right)
  + \left|\hat{\mathrm{S}}_{\bm{\lambda}}\right|\phi_n,
\end{equation}
where $\hat{\delta}\left(\bm{\lambda}\right) = \left(\hat{\bm{\mu}}^T, \hat{\bm{\beta}}^T\right)^T$ is obtained by solving (\ref{eq:penalized_objective_function}), and $\hat{\mathrm{S}}_{\bm{\lambda}}$ is the estimated model structure with $|\hat{\mathrm{S}}_{\bm{\lambda}}| = |\hat{\delta}(\bm{\lambda})| = \hat{K} + |\hat{\bm{\beta}}|_0$. In the definition, $\phi_n$ is a constant that can depend on the sample size $n$. We choose the tuning parameter $\bm{\lambda} = \left(\lambda_1, \lambda_2\right)$ by minimizing this modified BIC with the grid search.

\subsection{Practical tricks in computing}
\label{sec:techniques-practice}

In this subsection we discuss some computing tricks that can much speed up the algorithm.
\begin{enumerate}[({T}1)]
\item
  For the ADMM algorithm, we can stop earlier before its convergence   during the iterative steps  given tuning parameters $\lambda_1$ and $\lambda_2$. The ADMM algorithm can provide a fairly good result quickly at each step given a pair of tuning parameters which serves as the initial value of the next step in the solution path. The first several values of the tuning parameters $\lambda$ used in the solution path is usually considered as the warm start and will be discarded. It can significantly cut down the time consumption. In other words, we can set a quite loose tolerance for the ADMM algorithm.

\item
  Given the fact that this is an iterative algorithm, some clustering procedures can be used to convert the estimated intercept vector $\hat{\bm{\mu}}$ into a reasonable grouping result. When the algorithm meets a sharp tolerance, this procedure can be as simple as rounding on $\hat{\bm{\mu}}$ or hard-thresholding over $\left\{\hat{\mu}_i - \hat{\mu}_j\right\}_{i\neq j}$. When the tolerance is loosen, we use a simple K-means method to improve the grouping result. The number of centroids is selected according to the Average Silhouette Width \citep[ASW]{Kaufman1990p-}.

\end{enumerate}

 These tricks can efficiently improve the practical performance of the proposed algorithm. Furthermore,
  since most of the updating steps in the algorithm is implemented elementwisely, parallel  or distributed computations can be considered for large-scale datasets. 

\section{Asymptotic Properties}
\label{sec:asympt-properties}
In this section we give the asymptotic properties of the proposed estimator.  Under some regularity conditions, we show that the oracle estimator falls in the set of local minimizers of the object function with high probability. And the modified BIC provides asymptotic model selection consistency. First, we introduce some assumptions.

\begin{enumerate}[({A}1)]
\item\label{item:conditions_on_the_design}
  There exists a constant $M_1$ such that
  \[
    \left|x_{ij}\right|\leq M_1,\quad
    \forall 1 \leq i \leq n,\; 1\leq j\leq p,
  \]
  and there exist two positive constants $C_1$ and $C_2$ such that
  \[
    \begin{aligned}
      C_1 &\leq \lambda_{\mathrm{min}}\left(
      \frac{1}{n}
      \begin{pmatrix}
        \bm{Z} & \bm{X}_A
      \end{pmatrix}^T
      \begin{pmatrix}
        \bm{Z} & \bm{X}_A
      \end{pmatrix}
    \right)    \\
    & \leq
    \lambda_{\mathrm{max}}\left(
      \frac{1}{n}
      \begin{pmatrix}
        \bm{Z} & \bm{X}_A
      \end{pmatrix}^T
      \begin{pmatrix}
        \bm{Z} & \bm{X}_A
      \end{pmatrix}
    \right)
    \leq C_2
    ,
    \end{aligned}
    \]
    where $
    \begin{pmatrix}
      \bm{Z} & \bm{X}_A
    \end{pmatrix}
    $ is the covariate matrix under the true model, and $\lambda_{\mathrm{min}}\left(\cdot\right)$ and $\lambda_{\mathrm{max}}\left(\cdot\right)$ represent the operators to obtain the smallest and largest eigenvalues respectively.


  \item\label{item:conditions_on_true_model_dimension}
    $\mathrm{max}\left\{K_0, q\right\} = O\left(n^{c_1}\right)$ for some $0 \leq c_1 < \frac{1}{3}$.

  \item\label{item:conditions_on_smallest_siginal}
    There exist constants $c_2$ and $M_3$ such that
    \[
      2c_1 < c_2 \leq 1 \quad\text{and }\quad
      n^{\left(1 - c_2\right) / 2}  b_n \geq M_3,
    \]
    where $b_n = \mathrm{min}\left(
      \underset{i \neq j}{\mathrm{min}}\left|\alpha_{0, i} - \alpha_{0, j}\right|,
      \underset{1 \leq j \leq q}{\mathrm{min}}\left|\beta_{0, j}\right|
    \right)$.

  \item\label{item:conditions_on_random_error_1}
    There exists a positive constant $c_3$ such that for all constant $c\in\left[-c_3, c_3\right]$,
    \[
      P\left(
        \left|\psi\left(\varepsilon_i + c\right)\right|
        > x
      \right)
       \leq 2 \mathrm{exp}\left(-c_4 x ^ 2\right),
     \]
     where $c_4$ is some positive constant. 
   \item \label{item:conditions_on_random_error_2}
     $\mathrm{E}\psi\left(\varepsilon_i + \Delta\right)$ and $\mathrm{Var}\left(\psi\left(\varepsilon_i + \Delta\right)\right)$ are {uniformly continuous} in $\Delta$ with respect to the subjects $i$.
\end{enumerate}
Assumption (A\ref{item:conditions_on_the_design}) requires that the covariate is well behaved, which is also considered by \cite{ZHANG2019p-} and similar to (C1) of \cite{Ma2017p410-423} and \cite{wang1}.  The bounded condition for covariates is for the easy presentation of theoretical developments, and can be relaxed to $\left|x_{ij}\right|\leq M_1,\quad\forall 1 \leq i \leq n,\; 1\leq j\leq p$ with high probability. Assumption (A\ref{item:conditions_on_true_model_dimension}) allows the true model dimensions to increase with the sample size $n$, but in a slow rate. Assumption (A\ref{item:conditions_on_smallest_siginal}) imposes the restriction on the smallest signals, including the distances between different subgroups and the active coefficients. Assumption (A\ref{item:conditions_on_random_error_1}) is related with the score, which is quite mild for M-estimators. For the $L_2$ loss, $\psi\left(\varepsilon_i\right) = 2\varepsilon_i$, so this condition holds only if the error term itself is subgaussian. While for the $L_1$ and $Huber$ losses, their derivatives are bounded hence this assumption on error terms are satisfied immediately. {Assumption (A\ref{item:conditions_on_random_error_2}) imposes the restriction  on the random error and it generalizes Condition (C3) of \cite{wang1} which focuses on the $L_1$ loss.}

Under these conditions, we can prove that the oracle estimate lies in the set of local minimizers of the penalized objective function (\ref{eq:penalized_objective_function}).
\begin{thm}
  \label{thm:oracle_local_minimizer}
Under Assumptions (A\ref{item:conditions_on_the_design})--(A\ref{item:conditions_on_random_error_2}), if the following conditions $\mathrm{max}\left(\lambda_1, \lambda_2\right) = o\left(n ^{-\left(1 - c_2\right) / 2}\right)$, $\sqrt{q\left(K_0 + q\right)} = o\left(\sqrt{n}\lambda_2\right)$, $\left(K_0 + q\right)\mathrm{log}n = o\left(n\lambda_2\right)$, $n\lambda_2^2\to\infty$, $\mathrm{log}p = o\left(n\lambda_2^2\right)$ and $\sqrt{\mathrm{log}n}/(n\lambda_1\min\{\left|G_{k}\right|,k=1,\cdots,K_0\}) = o\left(1\right)$ hold, then with probability approaching to 1, the oracle estimate is among the local minimizers of the penalized regression function (\ref{eq:penalized_objective_function}) with the penalties SCAD or MCP.
\end{thm}

Next, we study the properties of the modified BIC, and mainly focus on the consistency of this criterion. Recall that the modified BIC is defined in (\ref{eq:modified_BIC}). And for the regular estimate obtained under a given parameter structure $\mathrm{S}$,
  \[
    \left(\tilde{\bm{\alpha}}_{\mathrm{S}}^T, \tilde{\bm{\beta}}_{\mathrm{S}}^T\right)^T =
    \underset{\bm{\alpha}, \bm{\beta}}{\mathrm{arg}\,\mathrm{min}}
    \frac{1}{n}\sum\limits_{i = 1}^n
    \rho\left(
      y_i - \bm{z}_{\mathrm{S}, i}^T\bm{\alpha} - \bm{x}_{\mathrm{S}, i}^T\bm{\beta}
    \right),
  \]
  where $\left\{\bm{z}_{\mathrm{S}, i}\right\}$ and $\left\{\bm{x}_{\mathrm{S}, i}\right\}$ are the grouping and covariate vectors corresponding to $\mathrm{S}$, we can also compute its modified BIC by
\begin{equation}
  \label{eq:oracle_modified_BIC}
  \mathrm{BIC}\left(\tilde{\delta}\left(\mathrm{S}\right)\right)
  = \mathrm{log}\left(
    \frac{1}{n}
    \sum\limits_{i = 1}^n\rho\left(
      y_i - \bm{z}_{\mathrm{S}, i}^T\tilde{\bm{\alpha}}_{\mathrm{S}} - \bm{x}_{\mathrm{S}, i}^T\tilde{\bm{\beta}}_{\mathrm{S}}
    \right)
  \right)
  + \left|\mathrm{S}\right|\phi_n
  ,
\end{equation}
where $\tilde{\delta}\left(\mathrm{S}\right) = \left(\tilde{\bm{\alpha}}_{\mathrm{S}}^T, \tilde{\bm{\beta}}_{\mathrm{S}}^T\right)^T$. Note that $\tilde{\delta}\left(\mathrm{S}_0\right)$ is the oracle estimate and we introduce some additional assumptions for establishing the selection consistency.
\par
Due to the sparsity, we shall focus on the model space $\mathcal{S}$ where the covariates $\bm{Z}$ and $\bm{X}$ are considered, and for any model $S \in \mathcal{S}$, the true number of subgroups $K_S$ and active covariates $q_S$ are bounded by $K_U$ and $q_U$, respectively, where $K_U \in \left(K_0, \infty\right)$ and $q_U \in \left(q_0, \infty\right)$, and $\mathrm{lim\;sup}_n\left(K_U + q_U\right) / n^{\kappa^\star} < 1$ for some $\kappa^\star < 1$. Also, for any submodel $S \in \mathcal{S}$, the matrix $\left(\bm{Z}_S, \bm{X}_S\right)$  satisfies Assumption (A\ref{item:conditions_on_the_design}) for the identifiability. Then we introduce some additional assumptions.
  \begin{enumerate}
  \item[(A6)] For any given model $S \in \mathcal{S}$, the classic estimate
    \[
      \left(\tilde{\bm{\alpha}}_{\mathrm{S}}^T, \tilde{\bm{\beta}}_{\mathrm{S}}^T\right)^T =
    \underset{\bm{\alpha}, \bm{\beta}}{\mathrm{arg}\,\mathrm{min}}
    \frac{1}{n}\sum\limits_{i = 1}^n
    \rho\left(
      y_i - \bm{z}_{\mathrm{S}, i}^T\bm{\alpha} - \bm{x}_{\mathrm{S}, i}^T\bm{\beta}
    \right)
    \] satisfies
    \[
      \left\|
        \left(\tilde{\bm{\alpha}}_{\mathrm{S}}^T, \tilde{\bm{\beta}}_{\mathrm{S}}^T\right)^T
        - \left(
          \bm{\alpha}_{\mathrm{S}}^T, \bm{\beta}_{\mathrm{S}}^T
        \right)^T
      \right\| = O_p\left(\frac{1}{\sqrt{n}}\right),
    \]
    where
    \[
      \left(\bm{\alpha}_{\mathrm{S}}^T, \bm{\beta}_{\mathrm{S}}^T\right)^T =
      \underset{\bm{\alpha}, \bm{\beta}}{\mathrm{arg}\,\mathrm{min}}\;
      \mathrm{E}\rho\left(
        Y - Z_{\mathrm{S}}^T\bm{\alpha} - X_{\mathrm{S}}^T\bm{\beta}
      \right),
    \]
    and the subscript $\mathrm{S}$ refers to that we only use those covariates $(Z_\mathrm{S},X_\mathrm{S})$ specified by the model $\mathrm{S}$.

  \item[(A7)] The classic M-estimate satisfies
    \[
      \begin{aligned}
        & \sum\limits_{i = 1}^n\rho\left(
          y_i
          - \bm{z}_{\mathrm{S},i}^T\tilde{\bm{\alpha}}_{\mathrm{S}}
          - \bm{x}_{\mathrm{S}, i}^T\tilde{\bm{\beta}}_{\mathrm{S}}
        \right)
        - \sum\limits_{i = 1}^n \rho\left(
          y_i
          - \bm{z}_{S,i}^T\bm{\alpha}_{\mathrm{S}}
          - \bm{x}_{S, i}^T\bm{\beta}_{\mathrm{S}}
        \right)    \\
        = & - \sum\limits_{i = 1}^n \psi\left(
          y_i
          - \bm{z}_{\mathrm{S},i}^T\bm{\alpha}_{\mathrm{S}}
          - \bm{x}_{\mathrm{S}, i}^T\bm{\beta}_{\mathrm{S}}
        \right)
        \left\{
          \bm{z}_{\mathrm{S}, i}^T\left(\tilde{\bm{\alpha}}_{\mathrm{S}} - \bm{\alpha}_{\mathrm{S}}\right)
          + \bm{x}_{\mathrm{S}, i}^T\left(\tilde{\bm{\beta}}_{\mathrm{S}} - \bm{\beta}_{\mathrm{S}}\right)
        \right\}
        + O_p\left(1\right)
        .
      \end{aligned}
    \]
  \end{enumerate}
  The classic results of Assumptions (A6)--(A7) have been established  under some mild conditions \citep{he2}. 
  Thus, we have the following theorem about the selection consistency.

\begin{thm}
  \label{thm:consistency_of_mBIC}
  Under Assumptions (A\ref{item:conditions_on_the_design}) -- (A7), for any sequence $\phi_n\to 0$ satisfying $\mathrm{log}\left(n + p\right) / n = o\left(\phi_n\right)$, we have
  \[
    P\left(
      \underset{\mathrm{S} \neq \mathrm{S}_0, S \in \mathcal{S}}{\mathrm{inf}}\;
      \mathrm{BIC}\left(\tilde{\delta}\left(\mathrm{S}\right)\right)
      > \mathrm{BIC}\left(\tilde{\delta}\left(\mathrm{S}_0\right)\right)
    \right)
    \to 1,
  \]
  where $\mathrm{S}_0$ refers to the true model.
\end{thm}

For any tuning parameter $\bm{\lambda} = \left(\lambda_1, \lambda_2\right)$, let $\hat{\mathrm{S}}_{\bm{\lambda}}$ denote the corresponding estimated model structure. By  (\ref{eq:modified_BIC}) and (\ref{eq:oracle_modified_BIC}), we have
\[
  \mathrm{BIC}\left(\hat{\delta}\left(\bm{\lambda}\right)\right)
  \geq \mathrm{BIC}\left(\tilde{\delta}\left(\hat{\mathrm{S}}_{\bm{\lambda}}\right)\right).
\]
By Theorem \ref{thm:oracle_local_minimizer} we know that with high probability, the true model structure $\mathrm{S}_0$ can be identified with the tuning parameter $\bm{\lambda}_0 = \left(\lambda_{01}, \lambda_{02}\right)$. Therefore $\mathrm{BIC}\left(\hat{\delta}\left(\bm{\lambda}_0\right)\right) = \mathrm{BIC}\left(\tilde{\delta}\left(\mathrm{S}_0\right)\right)$. Then by Theorem \ref{thm:consistency_of_mBIC}, we know that for any $\bm{\lambda}$ that leads to the wrong model structure, we have $\mathrm{BIC}\left(\hat{\delta}\left(\bm{\lambda}\right)\right) > \mathrm{BIC}\left(\hat{\delta}\left(\bm{\lambda}_0\right)\right)$ with the high probability.

\section{A Simulation Study}
\label{sec:simulation-result}
In this section we compare the performance of the proposed methods with its competitors under different scenarios.
\par
The data are  independently generated from the following model
\[
y_i = \mu_i + \bm{x}_i^T\bm{\beta} + \varepsilon_i,~~i = 1, 2, \cdots, n,
\]
 where $\bm{x}_i$ is the $p$-dimensional covariate, and $\varepsilon_i$ is the model error.  Under different scenarios, we will consider different combinations of the sample size $n$, the number of covariate $p$, the number of true signals $q$, the vector of treatment effects $\bm{\mu}$, the covariate effect $\bm{\beta}$, and the distributions of covariates and errors. In this paper, we only report the results of the competitors compared with proposed method with the SCAD penalty, and the performances of those with the penalty MCP are quite similar.

\par
We compare the following statistics for those methods considered in this simulation study.
\begin{enumerate}[(M1)]

\item
  MAE$_{\bm{\mu}}$ and MAE$_{\bm{\beta}}$: the mean absolute errors for the treatment and covariate effects respectively, given by
  \[
    \text{MAE}_{\bm{\mu}} = \frac{1}{n}\sum\limits_{i = 1}^n\left|\hat{\mu}_i - \mu_i\right|
    \text{\quad and \quad}
    \text{MAE}_{\bm{\beta}} = \frac{1}{p}\sum\limits_{i = 1}^p\left|\hat{\beta}_i - \beta_i\right|
    .
  \]

\item
  $\bar{K}$ and $\tilde{K}$: the average and median value of the estimated number of subgroups.

\item
  $\bar{q}$ and $\tilde{q}$: the average and median  of the estimated number of active covariates.

\item
  $RI$ \citep[Rand Index]{Rand_1971}: it is commonly used to describe how close two grouping results are. For a set of points $\left\{x_1, \cdots, x_n\right\}$ and two clustering results $C$ and $C'$, the RI for $C = \left\{c_1, \cdots, c_{K_1}\right\}$ and $C'_2 = \left\{c'_1, \cdots, c'_{K_2}\right\}$ is defined as
  \[
    RI\left(C, C'\right) = \frac{2}{n\left(n - 1\right)}
    \times
    \sum\limits_{1\leq i < j \leq n} \gamma_{ij}
    ,
  \]
  where
  \[
    \gamma_{ij} = \left\{
      \begin{aligned}
        & 1 && \text{$x_i$ and $x_j$ belong to the same subgroup in both $C$ and $C'$,}    \\
        & 1 && \text{$x_i$ and $x_j$ belong to the different subgroups in both $C$ and $C'$,}    \\
        & 0 && \text{otherwise}.
      \end{aligned}
    \right.
  \]
\end{enumerate}
The index $RI$ ranges from 0 to 1. The closer it is to one, the more similar these two clustering results are. When $RI = 1$, these two clustering results are identical. During the simulation, we compute the index $RI$ between the estimated subgroups and the real subgroup structures, and report the average and median values of $RI$ under each model setting.

\par
Throughout this simulation study, we consider the $L_1$, $L_2$ and $Huber$ losses in the proposed method for the comparisons with the method RSI \citep[Robust subgroup identification]{ZHANG2019p-}. These methods search over the same sequence of $\lambda_1$ and $\lambda_2$ whose upper bounds are given in Section \ref{sec:choice-lambdal-and-lambda2}. We choose the tuning parameters which minimize the modified BIC defined in (\ref{eq:modified_BIC}) with $\phi_n = c{\mathrm{log}n}\mathrm{log}\mathrm{log}\left(n + p\right)/{n}$. For the $L_2$ loss, $c$ is set to be 10 as suggested by \cite{Ma2017p410-423}. For the $L_1$ loss, $c$ is set to be 5, which is used by \cite{ZHANG2019p-}. For the $Huber$ loss, we also use $c = 5$. Meanwhile, the default  value $1.345$ is utilized for the $Huber$ loss.
\par
For the proposed method, we set the max number of iteration to be 50, which is a early stop of the iterations before the algorithm meets a sharp convergence criterion. The simulation study is conducted over 500 repeatedly generated Monte Carlo datasets under each simulation setting and we consider following scenarios in this simulation study.

\subsection{Subgroup analysis for low-dimensional cases}
\label{sec:moderate-n-small}

In this case, we consider $n = 200$ or $400$, $p=q=5$.  Since the number of considered predictors is small, the variable selection is no longer necessary. For subgroups, we consider $K=2$ with centers $\left\{-1, 1\right\}$ and $K = 3$ with centers $\left\{-2, 0, 2\right\}$. 
\par
More specifically, we use $\bm{x}_i = \left(x_{i, 1}, \cdots, x_{i, 5}\right)^T$ to represent the covariate and set the corresponding coefficient as $\bm{\beta} = \bm{1}_5$. The covariates and the intercepts are independently generated from the standard normal distribution and  the multinormial distribution, with equal probability at each group center, respectively. The errors $\varepsilon = 0.5 \epsilon$ are independently generated, where three different distributions are considered for the variable $\epsilon$: (i) the standard normal distribution $\mathrm{N}\left(0, 1\right)$; (ii) the t-distribution  with five degrees of freedom, denoted as $t\left(5\right)$; (iii) the Gaussian mixture $0.95\times \mathrm{N}\left(0, 1\right) + 0.05 \times \mathrm{N}\left(0, 10^2\right)$. Under these settings, approximately 2\% to 10\% of the data will be closer to other subgroup centers instead of its real center. 
\par
For these different settings of $n$, $K$ and error types, the results are summarized in Tables~\ref{tab:benchmark_n200_err1}--\ref{tab:benchmark_n200_err3}, respectively. 
In these tables, $L_1^{1000}$ represents the proposed method with the $L_1$ loss and up to 1000 iterations in the ADMM approximations. Although more iterations often lead to better results,  the improvement is relatively limited comparing to the one where up to 50 iterations are used.
For normally distributed errors, all methods perform well and comparably.
When the errors are generated from either t distribution or the mixture distribution, the proposed method with the loss $L_2$ is seriously affected by outliers, while the $L_1$ and $Huber$ losses provide much more robust results.

\begin{table}[htbp]
  \centering
  \footnotesize
  \begin{tabular}{r|r|rrrrrrr}
  \hline
 n  & K & Method          & MAE$_{\bm{\mu}}$ & MAE$_{\bm{\beta}}$ & $\bar{K}$ & $\tilde{K}$ & $\bar{RI}$ & $\tilde{RI}$ \\
    \hline
\multirow{10}{*}{200} & \multirow{5}{*}{2} &
           $L_1$          & 0.117(0.041) & 0.041(0.015) & 2.000(0.000) & 2 & 0.938(0.024) & 0.942 \\
    \cline{3-9}
    &   &  $L_2$          & 0.126(0.096) & 0.040(0.017) & 2.222(0.896) & 2 & 0.927(0.068) & 0.942 \\
    \cline{3-9}
    &   & $Huber$         & 0.110(0.055) & 0.038(0.015) & 1.998(0.045) & 2 & 0.940(0.030) & 0.942 \\
   \cline{3-9}
    &   &   RSI           & 0.122(0.082) & 0.064(0.046) & 2.004(0.063) & 2 & 0.940(0.037) & 0.942 \\
   \cline{3-9}
    &   & {$L_1^{1000}$}  & 0.136(0.071) & 0.045(0.019) & 2.146(0.511) & 2 & 0.924(0.052) & 0.932 \\

    \cline{2-9}
    \cline{2-9}
    & \multirow{5}{*}{3} &
           $L_1$          & 0.558(0.292) & 0.088(0.045) & 2.330(0.475) & 2 & 0.809(0.098) & 0.743 \\
    \cline{3-9}
    &   &  $L_2$          & 0.363(0.365) & 0.070(0.035) & 3.852(1.764) & 3 & 0.848(0.172) & 0.912 \\
    \cline{3-9}
    &   & $Huber$         & 0.233(0.164) & 0.063(0.033) & 3.102(0.537) & 3 & 0.912(0.054) & 0.926 \\
    \cline{3-9}
    &   &   RSI           & 0.276(0.263) & 0.102(0.084) & 2.868(0.413) & 3 & 0.907(0.080) & 0.937 \\
    \cline{3-9}
    &   & {$L_1^{1000}$}  & 0.243(0.121) & 0.071(0.038) & 3.316(0.798) & 3 & 0.903(0.049) & 0.914 \\
   \hline
   \hline
\multirow{10}{*}{400} & \multirow{5}{*}{2} &
           $L_1$          & 0.089(0.026) & 0.029(0.010) & 2.000(0.000) & 2 & 0.948(0.016) & 0.946 \\
    \cline{3-9}
    &   &  $L_2$          & 0.308(0.074) & 0.040(0.014) & 6.256(1.291) & 7 & 0.737(0.068) & 0.720 \\
    \cline{3-9}
    &   & $Huber$         & 0.085(0.025) & 0.026(0.009) & 2.000(0.000) & 2 & 0.948(0.016) & 0.946 \\
    \cline{3-9}
    &   &   RSI           & 0.107(0.135) & 0.050(0.073) & 2.068(0.687) & 2 & 0.943(0.049) & 0.951 \\
    \cline{3-9}
    &   & {$L_1^{1000}$}  & 0.100(0.031) & 0.031(0.012) & 2.072(0.259) & 2 & 0.938(0.022) & 0.942 \\
    \cline{2-9}
    \cline{2-9}
    & \multirow{5}{*}{3} &
           $L_1$          & 0.171(0.158) & 0.037(0.020) & 2.944(0.263) & 3 & 0.935(0.053) & 0.949 \\
    \cline{3-9}
    &   &  $L_2$          & 0.197(0.116) & 0.043(0.022) & 4.266(1.785) & 3 & 0.912(0.057) & 0.943 \\
    \cline{3-9}
    &   & $Huber$         & 0.128(0.049) & 0.034(0.014) & 3.036(0.207) & 3 & 0.946(0.019) & 0.947 \\
    \cline{3-9}
    &   &   RSI           & 0.184(0.278) & 0.087(0.144) & 3.060(0.661) & 3 & 0.938(0.059) & 0.949 \\
    \cline{3-9}
    &   & {$L_1^{1000}$}  & 0.542(0.105) & 0.062(0.023) & 3.280(0.753) & 3 & 0.788(0.026) & 0.792 \\
   \hline
\end{tabular}
  \caption{Low-dimensional case with Gaussian errors, and the standard deviations are reported in parentheses}
  \label{tab:benchmark_n200_err1}
\end{table}

\begin{table}[htbp]
  \centering
  \footnotesize
  \begin{tabular}{r|r|rrrrrrr}
  \hline
 n & K & Method& MAE$_{\bm{\mu}}$ & MAE$_{\bm{\beta}}$ & $\bar{K}$ & $\tilde{K}$ & $\bar{RI}$ & $\tilde{RI}$ \\
    \hline
\multirow{10}{*}{200} & \multirow{5}{*}{2} &
           $L_1$          & 0.172(0.044) & 0.045(0.017) & 2.000(0.000) & 2 & 0.890(0.031) & 0.896 \\
    \cline{3-9}
    &   &  $L_2$          & 0.473(0.385) & 0.056(0.026) & 1.944(1.226) & 2 & 0.748(0.186) & 0.861 \\
    \cline{3-9}
    &   & $Huber$         & 0.169(0.068) & 0.044(0.018) & 1.996(0.063) & 2 & 0.890(0.039) & 0.896 \\
    \cline{3-9}
    &   &   RSI           & 0.179(0.126) & 0.069(0.059) & 2.010(0.184) & 2 & 0.891(0.040) & 0.896 \\
    \cline{3-9}
    &   & {$L_1^{1000}$}  & 0.193(0.078) & 0.049(0.022) & 2.156(0.478) & 2 & 0.877(0.051) & 0.887 \\
    \cline{2-9}

    \cline{2-9}
    & \multirow{5}{*}{3} &
           $L_1$          & 0.734(0.155) & 0.105(0.042) & 2.078(0.283) & 2 & 0.747(0.049) & 0.735 \\
    \cline{3-9}
    &   &  $L_2$          & 0.738(0.501) & 0.087(0.038) & 3.504(2.326) & 3 & 0.672(0.246) & 0.810 \\
    \cline{3-9}
    &   & $Huber$         & 0.332(0.192) & 0.075(0.038) & 3.038(0.567) & 3 & 0.868(0.057) & 0.883 \\
    \cline{3-9}
    &   &   RSI           & 0.400(0.274) & 0.106(0.077) & 2.746(0.508) & 3 & 0.853(0.081) & 0.891 \\
    \cline{3-9}
    &   & {$L_1^{1000}$}  & 0.323(0.125) & 0.080(0.041) & 3.420(0.871) & 3 & 0.866(0.045) & 0.873 \\
    \hline
    \hline
\multirow{10}{*}{400} & \multirow{5}{*}{2} &
           $L_1$          & 0.146(0.029) & 0.031(0.011) & 2.000(0.000) & 2 & 0.899(0.019) & 0.900 \\
    \cline{3-9}
    &   &  $L_2$          & 0.365(0.180) & 0.041(0.015) & 4.916(1.969) & 6 & 0.752(0.099) & 0.742 \\
    \cline{3-9}
    &   & $Huber$         & 0.146(0.029) & 0.030(0.010) & 2.000(0.000) & 2 & 0.898(0.019) & 0.900 \\
    \cline{3-9}
    &   &   RSI           & 0.179(0.189) & 0.061(0.100) & 2.122(0.851) & 2 & 0.889(0.061) & 0.900\\
    \cline{3-9}
    &   & $L_1^{1000}$    & 0.157(0.036) & 0.033(0.013) & 2.110(0.313) & 2 & 0.890(0.024) & 0.891 \\
    \cline{2-9}

    \cline{2-9}
    & \multirow{5}{*}{3} &
           $L_1$          & 0.323(0.236) & 0.048(0.027) & 2.788(0.409) & 3 & 0.871(0.073) & 0.904 \\
    \cline{3-9}
    &   &  $L_2$          & 0.427(0.355) & 0.054(0.024) & 4.482(2.149) & 3 & 0.806(0.171) & 0.853 \\
    \cline{3-9}
    &   & $Huber$         & 0.219(0.106) & 0.042(0.018) & 3.002(0.233) & 3 & 0.901(0.032) & 0.906 \\
    \cline{3-9}
    &   &   RSI           & 0.302(0.333) & 0.103(0.171) & 3.080(0.975) & 3 & 0.885(0.069) & 0.906\\
    \cline{3-9}
    &   &   $L_1^{1000}$  & 0.597(0.110) & 0.067(0.024) & 3.056(0.744) & 3 & 0.772(0.026) & 0.775 \\
    \hline
\end{tabular}
  \caption{Low-dimensional case with the variable $\epsilon$ generated from the distribution $t(5)$, and the  standard deviations are reported in parentheses}
  \label{tab:benchmark_n200_err2}
\end{table}

\begin{table}[htbp]
  \centering
  \footnotesize
  \begin{tabular}{r|r|rrrrrrr}
  \hline
 n & K & Method& MAE$_{\bm{\mu}}$ & MAE$_{\bm{\beta}}$ & $\bar{K}$ & $\tilde{K}$ & $\bar{RI}$ & $\tilde{RI}$ \\
    \hline
\multirow{10}{*}{200} & \multirow{5}{*}{2} &
           $L_1$          & 0.196(0.080) & 0.049(0.019) & 2.094(0.292) & 2 & 0.883(0.037) & 0.887 \\
    \cline{3-9}
    &   &  $L_2$          & 0.995(0.124) & 0.088(0.035) & 1.196(0.500) & 1 & 0.509(0.057) & 0.501 \\
    \cline{3-9}
    &   & $Huber$         & 0.185(0.149) & 0.046(0.021) & 1.970(0.171) & 2 & 0.888(0.075) & 0.905 \\
   \cline{3-9}
    &   &   RSI           & 0.189(0.161) & 0.071(0.070) & 2.168(0.730) & 2 & 0.899(0.051) & 0.905 \\
   \cline{3-9}
    &   & {$L_1^{1000}$}  & 0.171(0.065) & 0.045(0.017) & 2.044(0.215) & 2 & 0.897(0.036) & 0.896 \\
   \cline{2-9}

    \cline{2-9}
    & \multirow{5}{*}{3} &
           $L_1$          & 0.782(0.047) & 0.108(0.041) & 2.002(0.045) & 2 & 0.725(0.015) & 0.725 \\
    \cline{3-9}
    &   &  $L_2$          & 1.380(0.183) & 0.120(0.044) & 1.152(0.781) & 1 & 0.351(0.095) & 0.333 \\
    \cline{3-9}
    &   & $Huber$         & 0.392(0.217) & 0.086(0.043) & 3.022(0.799) & 3 & 0.854(0.069) & 0.874 \\
   \cline{3-9}
    &   &   RSI           & 0.470(0.284) & 0.113(0.101) & 3.088(0.983) & 3 & 0.846(0.091) & 0.893 \\
   \cline{3-9}
    &   & {$L_1^{1000}$}  & 0.339(0.157) & 0.079(0.041) & 3.106(0.586) & 3 & 0.867(0.054) & 0.879 \\
   \hline
    \hline
\multirow{10}{*}{400} & \multirow{5}{*}{2} &
           $L_1$          & 0.161(0.057) & 0.032(0.011) & 2.172(0.393) & 2 & 0.903(0.023) & 0.902 \\
    \cline{3-9}
    &   &  $L_2$          & 1.013(0.093) & 0.060(0.022) & 1.332(0.539) & 1 & 0.503(0.041) & 0.499 \\
    \cline{3-9}
    &   & $Huber$         & 0.139(0.049) & 0.030(0.010) & 1.998(0.045) & 2 & 0.908(0.027) & 0.909 \\
    \cline{3-9}
    &   &   RSI           & 0.158(0.158) & 0.055(0.072) & 2.150(0.904) & 2 & 0.904(0.054) & 0.909\\
    \cline{3-9}
    &   &  $L_1^{1000}$   & 0.160(0.056) & 0.032(0.011) & 2.154(0.372) & 2 & 0.902(0.023) & 0.900 \\
    \cline{2-9}

    \cline{2-9}
    & \multirow{5}{*}{3} &
           $L_1$          & 0.743(0.145) & 0.070(0.028) & 2.060(0.238) & 2 & 0.733(0.050) & 0.722 \\
    \cline{3-9}
    &   &  $L_2$          & 1.377(0.167) & 0.080(0.028) & 1.138(0.590) & 1 & 0.346(0.082) & 0.333 \\
    \cline{3-9}
    &   & $Huber$         & 0.244(0.128) & 0.044(0.022) & 3.034(0.421) & 3 & 0.899(0.040) & 0.905 \\
    \cline{3-9}
    &   &   RSI           & 0.348(0.332) & 0.096(0.153) & 3.356(1.173) & 3 & 0.885(0.081) & 0.915\\
    \cline{3-9}
    &   & $L_1^{1000}$    & 0.736(0.089) & 0.068(0.027) & 2.254(0.508) & 2 & 0.731(0.022) & 0.724 \\
   \hline
\end{tabular}
  \caption{Low-dimensional case with the variable $\epsilon$ generated from the mixture distribution $0.95\times \mathrm{N}\left(0, 1\right) + 0.05 \times \mathrm{N}\left(0, 10^2\right)$, and the standard deviations are reported in parentheses}
  \label{tab:benchmark_n200_err3}
\end{table}

\par

\subsection{Subgroup analysis for high-dimensional cases}
\label{sec:moderate-n-large}

In this subsection, we consider $n \in\left\{200, 400\right\}$, $p\in\left\{50, 100\right\}$, $q=5$ and the covariate coefficient $\bm{\beta} = \left(\bm{1}_5^T, \bm{0}_{p - 5}^T\right)^T$. The subgroup intercepts $\mu_i$ follow the same multinormial distribution as described in the previous subsection, with $K\in\left\{2, 3\right\}$. The covariate matrix $\bm{X}$ is generated from the same distribution as that of Section \ref{sec:moderate-n-small}. We focus on distributions with thick tails of Section \ref{sec:moderate-n-small}. The results are reported in Tables~\ref{tab:rsavs_n200_p50_err2}--\ref{tab:rsavs_n200_p100_err3} for various settings.

\begin{table}[htbp]
\centering
\scriptsize
\begin{tabular}{r|r|rrrrrrrrr}
  \hline
 n & K & Method& MAE$_{\bm{\mu}}$ & MAE$_{\bm{\beta}}$ & $\bar{K}$ & $\tilde{K}$ & $\bar{q}$ & $\tilde{q}$ & $\bar{RI}$ & $\tilde{RI}$ \\
  \hline
\multirow{8}{*}{200} & \multirow{4}{*}{2} &
             $L_1$     & 0.220(0.070) & 0.006(0.005) & 2.000(0.000) & 2   & 4.990(0.161)  & 5  & 0.850(0.044) & 0.852   \\
    \cline{3-11}
    &   &    $L_2$     & 0.994(0.014) & 0.100(0.000) & 1.000(0.000) & 1   & 0.000(0.000)  & 0  & 0.501(0.000) & 0.501   \\
    \cline{3-11}
    &   &     $Huber$  & 0.289(0.236) & 0.009(0.008) & 1.970(0.371) & 2   & 5.322(0.805)  & 5  & 0.822(0.115) & 0.861   \\
    \cline{3-11}
    &   &       RSI    & 0.591(0.181) & 0.094(0.025) & 4.118(2.561) & 4   & 50.000(0.000) & 50 & 0.658(0.092) & 0.643   \\
    \cline{2-11}

    \cline{2-11}
    & \multirow{4}{*}{3} &
              $L_1$    & 0.812(0.080) & 0.018(0.013) & 2.000(0.000) & 2   & 4.788(0.572)  & 5  & 0.722(0.024) & 0.729   \\
    \cline{3-11}
    &   &     $L_2$    & 1.426(0.038) & 0.100(0.000) & 1.004(0.063) & 1   & 0.000(0.000)  & 0  & 0.333(0.004) & 0.333   \\
    \cline{3-11}
    &   &     $Huber$  & 0.626(0.278) & 0.019(0.016) & 2.486(0.520) & 2   & 5.058(1.112)  & 5  & 0.773(0.083) & 0.738   \\
    \cline{3-11}
    &   &       RSI    & 0.976(0.152) & 0.142(0.032) & 4.204(2.363) & 4   & 50.000(0.000) & 50 & 0.674(0.030) & 0.678 \\
   \hline
   \hline
\multirow{8}{*}{400} & \multirow{4}{*}{2} &
             $L_1$     & 0.158(0.034) & 0.003(0.001) & 2.000(0.000) & 2   & 5.000(0.000)  & 5  & 0.887(0.024) & 0.887   \\
   \cline{3-11}
    &   &    $L_2$     & 1.000(0.001) & 0.008(0.017) & 1.000(0.000) & 1   & 4.840(0.881)  & 5  & 0.499(0.000) & 0.499   \\
   \cline{3-11}
    &   &     $Huber$  & 0.169(0.068) & 0.004(0.003) & 2.020(0.189) & 2   & 5.006(0.077)  & 5  & 0.881(0.037) & 0.885   \\
   \cline{3-11}
    &   &       RSI    & 0.519(0.331) & 0.079(0.040) & 4.146(3.109) & 2   & 50.000(0.000) & 50 & 0.728(0.138) & 0.804 \\
   \cline{2-11}
   \cline{2-11}
    & \multirow{4}{*}{3} &
              $L_1$    & 0.771(0.024) & 0.009(0.003) & 2.000(0.000) & 2   & 5.016(0.126)  & 5  & 0.731(0.005) & 0.732   \\
   \cline{3-11}
    &   &     $L_2$    & 1.405(0.023) & 0.100(0.000) & 1.002(0.045) & 1   & 0.000(0.000)  & 0  & 0.333(0.001) & 0.333   \\
   \cline{3-11}
    &   &     $Huber$  & 0.349(0.227) & 0.008(0.004) & 2.820(0.385) & 3   & 5.544(1.224)  & 5  & 0.857(0.063) & 0.882   \\
   \cline{3-11}
    &   &       RSI    & 1.090(0.300) & 0.137(0.059) & 4.986(3.208) & 3   & 50.000(0.000) & 50 & 0.687(0.035) & 0.684 \\
   \hline
\end{tabular}
\caption{High-dimensional case with $p = 50$, $q = 5$, and the variable $\epsilon$ generated from the distribution $t(5)$, and the  standard deviations are reported in parentheses}
\label{tab:rsavs_n200_p50_err2}
\end{table}

\begin{table}[htbp]
\centering
\scriptsize
\begin{tabular}{r|r|rrrrrrrrr}
  \hline
 n & K & Method& MAE$_{\bm{\mu}}$ & MAE$_{\bm{\beta}}$ & $\bar{K}$ & $\tilde{K}$ & $\bar{q}$ & $\tilde{q}$ & $\bar{RI}$ & $\tilde{RI}$ \\
  \hline
\multirow{8}{*}{200} & \multirow{4}{*}{2} &
              $L_1$    & 0.270(0.119) & 0.008(0.008) & 2.028(0.165) & 2   & 4.956(0.300)  &  5 & 0.821(0.068) & 0.835   \\
    \cline{3-11}
    &   &     $L_2$    & 0.995(0.020) & 0.100(0.000) & 1.002(0.045) & 1   & 0.000(0.000)  &  0 & 0.501(0.001) & 0.501   \\
    \cline{3-11}
    &   &     $Huber$  & 0.770(0.322) & 0.014(0.014) & 1.372(0.496) & 1   & 4.962(0.864)  &  5 & 0.597(0.144) & 0.501   \\
    \cline{3-11}
    &   &       RSI    & 0.723(0.256) & 0.106(0.038) & 4.974(2.822) & 5   & 50.000(0.000) & 50 & 0.629(0.084) & 0.599 \\
    \cline{2-11}
    \cline{2-11}
    & \multirow{4}{*}{3} &
             $L_1$     & 0.824(0.079) & 0.018(0.013) & 2.000(0.000) & 2   & 4.804(0.578)  &  5 & 0.712(0.023) & 0.718   \\
    \cline{3-11}
    &   &    $L_2$     & 1.429(0.038) & 0.100(0.000) & 1.002(0.045) & 1   & 0.000(0.000)  &  0 & 0.333(0.000) & 0.333   \\
    \cline{3-11}
    &   &     $Huber$  & 0.818(0.281) & 0.026(0.018) & 2.174(0.541) & 2   & 4.694(1.095)  &  5 & 0.699(0.125) & 0.710   \\
    \cline{3-11}
    &   &       RSI    & 1.080(0.209) & 0.153(0.040) & 4.392(2.626) & 3.5 & 50.000(0.000) & 50 & 0.663(0.034) & 0.668 \\
   \hline
   \hline
\multirow{8}{*}{400} & \multirow{4}{*}{2} &
             $L_1$     & 0.179(0.053) & 0.004(0.002) & 2.068(0.252) & 2   & 5.000(0.000)  &  5 & 0.881(0.032) & 0.883   \\
   \cline{3-11}
    &   &    $L_2$     & 1.001(0.012) & 0.091(0.027) & 1.016(0.126) & 1   & 0.468(1.447)  &  0 & 0.499(0.000) & 0.499   \\
   \cline{3-11}
    &   &     $Huber$  & 0.505(0.385) & 0.006(0.004) & 1.630(0.487) & 2   & 5.408(1.152)  &  5 & 0.721(0.176) & 0.808   \\
   \cline{3-11}
    &   &       RSI    & 0.548(0.344) & 0.074(0.034) & 4.850(3.415) & 2   & 50.000(0.000) & 50 & 0.718(0.135) & 0.792 \\
   \cline{2-11}
   \cline{2-11}
    & \multirow{4}{*}{3} &
              $L_1$    & 0.781(0.028) & 0.009(0.003) & 2.000(0.000) & 2   & 5.028(0.188)  &  5 & 0.721(0.007) & 0.722   \\
   \cline{3-11}
    &   &     $L_2$    & 1.408(0.025) & 0.100(0.000) & 1.002(0.045) & 1   & 0.002(0.045)  &  0 & 0.333(0.000) & 0.333   \\
   \cline{3-11}
    &   &     $Huber$  & 0.548(0.302) & 0.010(0.005) & 2.500(0.532) & 3   & 5.338(0.922)  &  5 & 0.801(0.095) & 0.735   \\
   \cline{3-11}
    &   &       RSI    & 1.136(0.315) & 0.136(0.057) & 5.668(3.461) & 6   & 50.000(0.000) & 50 & 0.684(0.030) & 0.684 \\
   \hline
\end{tabular}
\caption{High-dimensional case with $p = 50$, $q = 5$, and the variable $\epsilon$ generated from the mixture distribution $0.95\times \mathrm{N}\left(0, 1\right) + 0.05 \times \mathrm{N}\left(0, 10^2\right)$, and the standard deviations are reported in parentheses}
\label{tab:rsavs_n200_p50_err3}
\end{table}

\begin{table}[htbp]
\centering
\scriptsize
\begin{tabular}{r|r|rrrrrrrrr}
  \hline
 n & K & Method& MAE$_{\bm{\mu}}$ & MAE$_{\bm{\beta}}$ & $\bar{K}$ & $\tilde{K}$ & $\bar{q}$ & $\tilde{q}$ & $\bar{RI}$ & $\tilde{RI}$ \\
  \hline
\multirow{8}{*}{200} & \multirow{4}{*}{2} &
             $L_1$     & 0.232(0.066) & 0.003(0.002) & 2.000(0.000) & 2 & 5.000(0.063)   &   5 & 0.842(0.044) & 0.844     \\
   \cline{3-11}
    &   &    $L_2$     & 0.994(0.013) & 0.050(0.000) & 1.000(0.000) & 1 & 0.000(0.000)   &   0 & 0.501(0.000) & 0.501     \\
   \cline{3-11}
    &   &     $Huber$  & 0.371(0.286) & 0.005(0.005) & 1.922(0.461) & 2 & 5.418(1.007)   &   5 & 0.779(0.135) & 0.844     \\
   \cline{3-11}
    &   &       RSI    & 0.893(0.231) & 0.123(0.035) & 6.398(2.666) & 6 & 100.000(0.000) & 100 & 0.535(0.024) & 0.532 \\
   \cline{2-11}
   \cline{2-11}
    & \multirow{4}{*}{3} &
             $L_1$     & 0.817(0.084) & 0.009(0.007) & 2.000(0.000) & 2 &   4.780(0.670) &   5 & 0.720(0.026) & 0.726     \\
   \cline{3-11}
    &   &    $L_2$     & 1.425(0.036) & 0.050(0.000) & 1.004(0.063) & 1 &   0.000(0.000) &   0 & 0.333(0.005) & 0.333     \\
   \cline{3-11}
    &   &     $Huber$  & 0.720(0.261) & 0.011(0.008) & 2.314(0.494) & 2 &   4.844(1.011) &   5 & 0.742(0.087) & 0.728     \\
   \cline{3-11}
    &   &       RSI    & 1.325(0.302) & 0.186(0.050) & 6.470(2.599) & 7 & 100.000(0.000) & 100 & 0.610(0.043) & 0.623 \\
   \hline
   \hline
\multirow{8}{*}{400} & \multirow{4}{*}{2} &
             $L_1$     & 0.161(0.035) & 0.002(0.001) & 2.000(0.000) & 2 &   5.000(0.000) &   5 & 0.883(0.024) & 0.883     \\
    \cline{3-11}
    &   &    $L_2$     & 1.000(0.001) & 0.004(0.010) & 1.000(0.000) & 1 &   4.790(1.004) &   5 & 0.499(0.000) & 0.499     \\
    \cline{3-11}
    &   &     $Huber$  & 0.179(0.076) & 0.002(0.001) & 2.018(0.133) & 2 &   5.004(0.063) &   5 & 0.873(0.037) & 0.878     \\
    \cline{3-11}
    &   &       RSI    & 0.853(0.349) & 0.093(0.041) & 6.976(2.929) & 8 & 100.000(0.000) & 100 & 0.578(0.079) & 0.547     \\
    \cline{2-11}
    \cline{2-11}
    & \multirow{4}{*}{3} &
              $L_1$    & 0.772(0.025) & 0.004(0.002) & 2.000(0.000) & 2 &   5.036(0.186) &   5 & 0.731(0.005) & 0.731     \\
    \cline{3-11}
    &   &     $L_2$    & 1.406(0.026) & 0.050(0.000) & 1.006(0.100) & 1 &   0.000(0.000) &   0 & 0.333(0.004) & 0.333     \\
    \cline{3-11}
    &   &     $Huber$  & 0.439(0.273) & 0.004(0.002) & 2.672(0.470) & 3 &   5.350(0.981) &   5 & 0.832(0.075) & 0.871     \\
    \cline{3-11}
    &   &       RSI    & 1.321(0.512) & 0.145(0.059) & 7.500(2.651) & 8.5 & 100.000(0.000) & 100 & 0.655(0.025) & 0.657 \\
   \hline
\end{tabular}
\caption{High-dimensional case with $p = 100$, $q = 5$, and the variable $\epsilon$ generated from the distribution $t(5)$, and the  standard deviations are reported in parentheses}
\label{tab:rsavs_n200_p100_err2}
\end{table}

\begin{table}[htbp]
\centering
\scriptsize
\begin{tabular}{r|r|rrrrrrrrr}
  \hline
 n & K & Method& MAE$_{\bm{\mu}}$ & MAE$_{\bm{\beta}}$ & $\bar{K}$ & $\tilde{K}$ & $\bar{q}$ & $\tilde{q}$ & $\bar{RI}$ & $\tilde{RI}$ \\
  \hline
\multirow{8}{*}{200} & \multirow{4}{*}{2} &
             $L_1$     & 0.277(0.106) & 0.004(0.004) & 2.014(0.118) & 2 &   4.956(0.300) &   5 & 0.814(0.066) & 0.819     \\
   \cline{3-11}
    &   &    $L_2$     & 0.994(0.022) & 0.050(0.000) & 1.002(0.045) & 1 &   0.000(0.000) &   0 & 0.501(0.000) & 0.501     \\
   \cline{3-11}
    &   &     $Huber$  & 0.807(0.288) & 0.008(0.007) & 1.360(0.051) & 1 &   4.910(0.867) &   5 & 0.579(0.124) & 0.501     \\
   \cline{3-11}
    &   &       RSI    & 1.030(0.268) & 0.139(0.044) & 6.152(2.737) & 6 & 100.000(0.000) & 100 & 0.529(0.020) & 0.526     \\
   \cline{2-11}
   \cline{2-11}
    & \multirow{4}{*}{3} &
             $L_1$     & 0.835(0.086) & 0.010(0.007) & 1.998(0.045) & 2 &   4.788(0.657) &   5 & 0.709(0.030) & 0.716     \\
   \cline{3-11}
    &   &    $L_2$     & 1.429(0.040) & 0.050(0.000) & 1.004(0.063) & 1 &   0.000(0.000) &   0 & 0.333(0.005) & 0.333     \\
   \cline{3-11}
    &   &    $Huber$   & 0.885(0.244) & 0.013(0.009) & 2.040(0.493) & 2 &   4.562(1.014) &   5 & 0.673(0.125) & 0.699     \\
   \cline{3-11}
    &   &       RSI    & 1.448(0.356) & 0.202(0.057) & 6.454(2.710) & 6 & 100.000(0.000) & 100 & 0.599(0.050) & 0.616     \\
    \hline
    \hline
\multirow{8}{*}{400} & \multirow{4}{*}{2} &
             $L_1$     & 0.180(0.063) & 0.002(0.001) & 2.048(0.232) & 2 &   5.004(0.089) &   5 & 0.877(0.037) & 0.883     \\
    \cline{3-11}
    &   &    $L_2$     & 1.000(0.006) & 0.047(0.011) & 1.004(0.063) & 1 &   0.308(1.190) &   0 & 0.499(0.000) & 0.499     \\
    \cline{3-11}
    &   &     $Huber$  & 0.592(0.392) & 0.003(0.002) & 1.524(0.500) & 2 &   5.306(0.985) &   5 & 0.679(0.176) & 0.741     \\
    \cline{3-11}
    &   &       RSI    & 0.901(0.342) & 0.089(0.040) & 8.032(2.503) & 9 & 100.000(0.000) & 100 & 0.572(0.059) & 0.560      \\
    \cline{2-11}
    \cline{2-11}
    & \multirow{4}{*}{3} &
              $L_1$    & 0.783(0.028) & 0.004(0.002) & 2.000(0.000) & 2 &   5.092(0.316) &   5 & 0.720(0.008) & 0.720     \\
    \cline{3-11}
    &   &     $L_2$    & 1.408(0.024) & 0.050(0.000) & 1.000(0.000) & 1 &   0.000(0.000) &   0 & 0.333(0.000) & 0.333     \\
    \cline{3-11}
    &   &     $Huber$  & 0.635(0.290) & 0.006(0.002) & 2.356(0.504) & 2 &   5.316(0.900) &   5 & 0.774(0.092) & 0.724     \\
    \cline{3-11}
    &   &       RSI    & 1.371(0.491) & 0.144(0.057) & 7.800(2.554) & 9 & 100.000(0.000) & 100 & 0.655(0.023) & 0.655     \\
    \hline
\end{tabular}
\caption{High-dimensional case with $p = 100$, $q = 5$, and the variable $\epsilon$ generated from the mixture distribution $0.95\times \mathrm{N}\left(0, 1\right) + 0.05 \times \mathrm{N}\left(0, 10^2\right)$, and the standard deviations are reported in parentheses}
\label{tab:rsavs_n200_p100_err3}
\end{table}

\par
As demonstrated in the simulation results, the method with the $L_2$ loss performs poorly since it can hardly tolerate outliers.  Also, it fails to recover the active covariates probably because the constant in mBIC is set to be 10, which strengthens the penalty effects compared with its competitors. The proposed method equipped with either the $L_1$ or the $Huber$ losses perform well, which can effectively identify both the grouping structure and  the active covariates. 
In the high-dimensional cases, the performance of the method RSI is seriously affected by the size of covariate $p$ due to the lack of variable selection procedures.
\par

\subsection{Subgroup analysis for relatively large data}
\label{sec:large-n-with}

When using the pairwise penalty method for the subgroup analysis, the algorithm has to deal with the paired treatment effects, which is quite challenging when the sample size $n$ is not small. \cite{ZHANG2019p-} pointed out that the method RSI would have to take a divide-and-conquer strategy when $n$ is large. At the dividing step, the method RSI is performed on each batch of data. At the conquering step, another procedure RSI is performed on the unique intercepts $\hat{\mu}_i$ gathered from each batch. However, our method is well ready for data stored in batches. In the following study, the batch size is set to be 200.

\par
For our method, the update of parameters $\bm{z}$, $\bm{s}$ and $\bm{w}$ is carried out elementwisely, which implies that we can implement parallel computing that can potentially deal with large-scale datasets. 

\par
In this subsection, we let $n = 1000$ and $K\in\left\{2, 3\right\}$. The group centers are $\left\{-2, 0, 2\right\}$ when $K = 3$ and $\left\{-1, 1\right\}$ for $K = 2$. For the covariate effect, we set $p = q = 5$ and $\bm{\beta}_0 = \bm{1}_q $. The predictors and the individual intercepts are generated in the same manner as described in previous subsections. The results are summarized in Tables~\ref{tab:largeN_err1}--\ref{tab:largeN_err3} for those three error distributions considered in the previous subsection respectively.

\begin{table}[htbp]
\centering
\footnotesize
\begin{tabular}{r|rrrrrrr}
  \hline
 K & Method& MAE$_{\bm{\mu}}$ & MAE$_{\bm{\beta}}$ & $\bar{K}$ & $\tilde{K}$ & $\bar{RI}$ & $\tilde{RI}$ \\
  \hline
\multirow{4}{*}{2} &
         $L_1$     & 0.073(0.015) & 0.018(0.006) & 2.000(0.000) & 2 & 0.952(0.010) & 0.951   \\
  \cline{2-8}
    &    $L_2$     & 0.291(0.038) & 0.025(0.009) & 6.732(0.537) & 7 & 0.734(0.036) & 0.729   \\
  \cline{2-8}
    &     $Huber$  & 0.069(0.015) & 0.016(0.005) & 2.000(0.000) & 2 & 0.952(0.010) & 0.951   \\
  \cline{2-8}
    &     RSI      & 0.688(0.375) & 0.305(0.176) & 6.418(2.690) & 6 & 0.699(0.101) & 0.685   \\
  \hline
  \hline
\multirow{4}{*}{3} &
         $L_1$     & 0.100(0.022) & 0.020(0.007) & 3.008(0.089) & 3 & 0.955(0.009) & 0.955   \\
  \cline{2-8}
    &    $L_2$     & 0.276(0.036) & 0.036(0.013) & 6.964(0.339) & 7 & 0.854(0.017) & 0.853   \\
  \cline{2-8}
    &     $Huber$  & 0.094(0.018) & 0.020(0.007) & 3.002(0.045) & 3 & 0.956(0.008) & 0.956   \\
  \cline{2-8}
    &     RSI      & 0.729(0.337) & 0.297(0.175) & 6.448(2.492) & 6 & 0.755(0.066) & 0.750   \\
  \hline
\end{tabular}
\caption{Relative large-scale case with Gaussian errors, and the  standard deviations are reported in parentheses}
\label{tab:largeN_err1}
\end{table}

\begin{table}[htbp]
\centering
\footnotesize
\begin{tabular}{r|rrrrrrr}
  \hline
 K & Method& MAE$_{\bm{\mu}}$ & MAE$_{\bm{\beta}}$ & $\bar{K}$ & $\tilde{K}$ & $\bar{RI}$ & $\tilde{RI}$ \\
  \hline
\multirow{4}{*}{2} &
         $L_1$     & 0.129(0.019) & 0.019(0.006) & 2.000(0.000) & 2   & 0.901(0.013) & 0.900   \\
  \cline{2-8}
    &    $L_2$     & 0.375(0.046) & 0.029(0.009) & 6.734(0.660) & 7   & 0.706(0.031) & 0.704   \\
  \cline{2-8}
    &     $Huber$  & 0.129(0.018) & 0.018(0.006) & 2.000(0.000) & 2   & 0.901(0.013) & 0.900   \\
  \cline{2-8}
    &     RSI      & 0.723(0.376) & 0.304(0.181) & 6.594(2.761) & 7   & 0.677(0.090) & 0.667   \\
  \hline
  \hline
\multirow{4}{*}{3} &
         $L_1$     & 0.177(0.036) & 0.021(0.008) & 2.990(0.045) & 3   & 0.911(0.013) & 0.912   \\
  \cline{2-8}
    &    $L_2$     & 0.371(0.057) & 0.039(0.014) & 6.972(0.368) & 7   & 0.821(0.026) & 0.820   \\
  \cline{2-8}
    &     $Huber$  & 0.171(0.022) & 0.022(0.008) & 3.000(0.000) & 3   & 0.912(0.010) & 0.912   \\
  \cline{2-8}
    &     RSI      & 0.789(0.338) & 0.280(0.172) & 6.400(2.444) & 6   & 0.732(0.056) & 0.726   \\
  \hline
\end{tabular}
\caption{Relative large-scale case with the variable $\epsilon$ generated from the distribution $t(5)$, and the standard deviations are reported in parentheses}
\label{tab:largeN_err2}
\end{table}

\begin{table}[htbp]
\centering
\footnotesize
\begin{tabular}{r|rrrrrrr}
  \hline
 K & Method& MAE$_{\bm{\mu}}$ & MAE$_{\bm{\beta}}$ & $\bar{K}$ & $\tilde{K}$ & $\bar{RI}$ & $\tilde{RI}$ \\
  \hline
\multirow{4}{*}{2} &
         $L_1$     & 0.128(0.032) & 0.020(0.007) & 2.072(0.266) & 2 & 0.913(0.013) & 0.914   \\
  \cline{2-8}
    &    $L_2$     & 1.020(0.032) & 0.038(0.013) & 1.530(0.784) & 1 & 0.500(0.000) & 0.500   \\
  \cline{2-8}
    &     $Huber$  & 0.123(0.019) & 0.018(0.006) & 2.000(0.000) & 2 & 0.915(0.013) & 0.916   \\
  \cline{2-8}
    &     RSI      & 0.784(0.411) & 0.310(0.187) & 7.666(3.277) & 8 & 0.680(0.096) & 0.669   \\
  \hline
  \hline
\multirow{4}{*}{3} &
         $L_1$     & 0.739(0.169) & 0.044(0.017) & 2.090(0.326) & 2 & 0.732(0.058) & 0.715   \\
  \cline{2-8}
    &    $L_2$     & 1.374(0.058) & 0.050(0.017) & 1.178(0.532) & 1 & 0.336(0.030) & 0.333   \\
  \cline{2-8}
    &     $Huber$  & 0.179(0.025) & 0.024(0.009) & 3.008(0.089) & 3 & 0.919(0.010) & 0.919   \\
  \cline{2-8}
    &     RSI      & 0.824(0.330) & 0.282(0.177) & 7.854(2.758) & 7 & 0.735(0.056) & 0.731   \\
  \hline
\end{tabular}
\caption{Relative large-scale cases with  the variable $\epsilon$ generated from the mixture distribution $0.95\times \mathrm{N}\left(0, 1\right) + 0.05 \times \mathrm{N}\left(0, 10^2\right)$, and the standard deviations are reported in parentheses}
\label{tab:largeN_err3}
\end{table}

\par
In general, our method performs well with those robust losses. The $L_1$ loss may lead to smaller estimates of the number of subgroups than that of the $Huber$ loss, while it is challenging for the method RSI coupled with divide-and-conquer strategy to identify the true subgroup structure. It is not surprising because  the batch size is limited and the communications between batches are insufficient for the divide-and-conquer strategy, especially when the real number of subgroups is not small. This implies the divide-and-conquer strategy should be carefully used in subgroup analysis.

\section{Conclusion}
In this paper, we have proposed a penalized method that can potentially deal with large-scale data in subgroup analysis. An approximation algorithm is carefully designed by adopting the spirit of the ADMM method, so the proposed M-estimator can take into account the high-dimensional data with outliers, and its estimation procedure can be implemented in a parallel or even distributed manner. As high-dimensional data are considered in subgroup analysis, theoretical development appears to be challenging, and few works have been done to account for such scenarios to our best knowledge.

As the sample size is small or moderate, we can similarly consider the local linear approximation in subgroup analysis of \cite{ZHANG2019p-} to improve its practical computing efficiency. However, for large-scale data, the divide-and-conquer strategy should be carefully used because sufficient communications  among those data subsets are probably necessary to provide a meaningful clustering result. Furthermore, it is still challenging to well identify true grouping structures  when there are too many real subgroups, which is intrinsically difficult, and the BIC tends to merge subgroups so more efforts are still needed to propose  better  criteria of choosing tuning parameters.


\clearpage
\begin{center}
{\large\bf SUPPLEMENTARY MATERIAL}
\end{center}

\begin{description}

\item[Online supplementary material] Supplementary Material is available online which includes details of the development of the proposed algorithm and the proofs of Theorems 1--3.
\end{description}

\bibliographystyle{asa}
\bibliography{lib}




\end{document}